\documentclass[aps,prb,superscriptaddress,reprint,floatfix]{revtex4-2}

\usepackage{float}
\usepackage{amsmath}
\usepackage{graphicx}
\usepackage{longtable}
\usepackage{colortbl}
\usepackage[english]{babel}
\usepackage{array}
\usepackage{multirow}
\usepackage{tabularx}
\usepackage{booktabs}
\usepackage{bm}

\usepackage{svg} 
\begin{document}

\title{Isotope effects on molecular structures and electronic properties of liquid water via deep potential molecular dynamics based on SCAN functional}

\author{Jianhang Xu}
\email{jianhang.xu@temple.edu}
\affiliation{Department of Physics, Temple University, Philadelphia, Pennsylvania 19122, USA}
\author{Chunyi Zhang}
\email{zhangchunyi.anna@gmail.com}
\affiliation{Department of Physics, Temple University, Philadelphia, Pennsylvania 19122, USA}
\author{Linfeng Zhang}
\affiliation{Program in Applied and Computational Mathematics, Princeton University, Princeton, New Jersey 08544, USA}
\author{Mohan Chen}
\affiliation{CAPT, HEDPS, College of Engineering, Peking University, Beijing 100871, China}
\author{Biswajit Santra}
\affiliation{Department of Physics, Temple University, Philadelphia, Pennsylvania 19122, USA}
\author{Xifan Wu}
\affiliation{Department of Physics, Temple University, Philadelphia, Pennsylvania 19122, USA}
\affiliation{Institute for Computational Molecular Science, Temple University, Philadelphia, Pennsylvania 19122, USA}


\begin{abstract}

Feynman path-integral deep potential molecular dynamics (PI-DPMD) calculations have been employed to study both light (H$_2$O) and heavy water (D$_2$O) within the isothermal–isobaric ensemble.
In particular, the deep neural network is trained based on \textit{ab initio} data obtained from the strongly constrained and appropriately normed (SCAN) exchange-correlation functional. 
Because of the lighter mass of hydrogen than deuteron, the properties of light water is more influenced by nuclear quantum effect than those of heavy water. 
Clear isotope effects are observed and analyzed in terms of hydrogen-bond structure and electronic properties of water that are closely associated with experimental observables. 
The molecular structures of both liquid H$_2$O and D$_2$O agree well with the data extracted from  scattering experiments. 
The delicate isotope effects on radial distribution functions and angular distribution functions are well reproduced as well. 
Our approach demonstrates that deep neural network combined with SCAN functional based \textit{ab initio} molecular dynamics provides an accurate theoretical tool for modeling water and its isotope effects.

\end{abstract}

\maketitle

\section{INTRODUCTION}

\par 
Liquid water as one of the most important chemical substances on earth has significant impacts on various chemical and biological processes \cite{ball_water_2008}.
Despite its simple structure as a three-atom molecule, water is of great complexity in its liquid condensed phase, where water molecules are hydrogen(H)-bonded with each other and form nearly tetrahedral networks \cite{eisenberg_structure_2005}.
H-bond is much weaker than the O-H covalent bond and constantly breaks and reforms under thermal fluctuations at the time scale of picoseconds \cite{zhang_first_2011}, which is hard to capture precisely in experiments.
Moreover, due to the light mass of hydrogen atoms, nuclear quantum effects (NQEs) induce nontrivial impacts on the H-bond network \cite{morrone_nuclear_2008,ceriotti_nuclear_2016,soper_quantum_2008}.
Therefore, fundamental experimental and theoretical studies on the structure of water have been at the center of scientific interests for decades \cite{thamer_ultrafast_2015,kim_maxima_2017,chen_hydroxide_2018,soper_is_2019}.

\par 
In experiment, various advanced techniques, such as x-ray and neutron diffraction \cite{soper_radial_2013,skinner_benchmark_2013}, infrared and Raman spectroscopy \cite{brubach_signatures_2005,auer_hydrogen_2007}, x-ray absorption and photoemission spectroscopy \cite{fransson_x-ray_2016,winter_photoemission_2006}, have been employed to study liquid water.
X-ray and neutron diffraction experiments are widely adopted to probe the time-averaged structural information of liquid water.
Combined with post-processing techniques, such as the empirical potential structure refinement (EPSR) method \cite{soper_tests_2001,soper_partial_2005}, the diffraction cross section data can be converted to commonly used structural properties, i.e. the radial distribution functions (RDFs) and angular distribution functions (ADFs).
The infrared and Raman spectroscopy are used to probe vibration modes and dynamics of water \cite{brubach_signatures_2005,auer_hydrogen_2007}.
Whereas the x-ray absorption and photoemission spectroscopy provide electronic structure information, which is directly related to local bonding configurations at molecular scale \cite{fransson_x-ray_2016,winter_photoemission_2006}.
However, the interpretation of experimental results in terms of structure is not straightforward and may induce artificial errors \cite{ceriotti_nuclear_2016}, making \textit{ab initio} molecular dynamics (AIMD) \cite{car_unified_1985} simulation an important theoretical tool to provide microscopic structural information of water on a sub-picosecond time scale.
In AIMD, the forces acting on atoms are generated ``on the fly" by the instantaneous first-principle based electronic ground state of the atomic structure without any empirical input.
AIMD can be used to predict structural, electronic, and dynamic properties of a given system at certain thermodynamic conditions.
Since the early simulation of water in the 1990s \cite{laasonen_water_1992,tuckerman_ab_1995}, AIMD has become an ideal approach to explore various chemical and biological aqueous systems.

\par 
In AIMD simulations, density functional theory (DFT) \cite{hohenberg_inhomogeneous_1964,kohn_self-consistent_1965} is adopted to calculate the potential energy surface (PES) at each time step due to a reasonable balance between accuracy and computational cost.
Even though DFT is an exact theory in principle, the practical form of exchange-correlation (XC) functional requires approximations. 
Different levels of approximations for the XC term have been developed to increase the predictive power of DFT, including the local density approximation (LDA), generalized gradient approximation (GGA), meta-GGA, and hybrid functionals \cite{perdew_jacobs_2001}.
Early studies on water clusters \cite{laasonen_water_1992,laasonen_structures_1993} showed that the LDA approximation overestimates H-bond strengths and leads to excessive short distances between water molecules.
The widely adopted GGA-level XC functionals, such as Perdew-Burke-Ernzerhof (PBE) \cite{perdew_generalized_1996,perdew_generalized_1996}, improve the structure of water, but still yield discrepancies with experimental observations \cite{gillan_perspective_2016}. 
The remaining problems of GGA functionals can be summarized in two key points.
For one, conventional XC functionals at the GGA level do not capture intermediate- and long-range non-local van der Waals (vdW) interactions \cite{hermann_first-principles_2017}, which have been proved to be crucial for the correct prediction of water density \cite{gaiduk_density_2015}.
Secondly, the well-known self-interaction error \cite{perdew_self-interaction_1981} of GGA functionals leads to the delocalization of protons in liquid water, which increases the stiffness of H-bond artificially.
The above drawbacks can be largely compensated by mixing a fraction of the exact exchange and adding vdW corrections \cite{miceli_isobaric_2015,wiktor_note:_2017}.
For example, studies adopting PBE0 hybrid functional and Tkatchenko Schefﬂer \cite{tkatchenko_accurate_2009} dispersion corrections (PBE0+TS) have shown improved covalent bond vibration frequencies and less overstructured RDF and ADF distributions \cite{distasio_individual_2014,ko_isotope_2019}. 
However, the application of such AIMD method depends semi-empirical treatments of vdW corrections and the hybrid functional requires a significant amount of computational cost.
The strongly constrained and appropriately normed (SCAN) meta-GGA functional \cite{sun_strongly_2015}, which satisfies all 17 known exact constraints on the semilocal XC functional and includes intermediate-range vdW interactions, is a viable option to address the above issues. 
Recent works have proved that the SCAN functional provides accurate descriptions of O-H covalent bond and H-bonding strength as well as dynamical properties in water clusters, liquid water, and ice with the level of accuracy comparable to  vdW inclusive hybrid functional \cite{sun_accurate_2016,chen_ab_2017,zheng_structural_2018,xu_first-principles_2019,sharkas_self-interaction_2020}. 
Moreover the computational cost of SCAN functional is marginally higher than GGA but about an order of magnitude less than hybrid functionals when applied to condensed phases \cite{furness_accurate_2020}.

\par 
Besides the proper choice of XC functionals, NQEs are indispensable for any water model to have quantitative agreement with the experimental observations as well.
Experimentally, NQEs can be identified in the water system by isotope substitution of hydrogen by deuterium.
In the joint x-ray/neutron diffraction experiments by Soper \textit{et al.} \cite{soper_quantum_2008}, distinct isotope effects, such as covalent bond contractions from H$_2$O to D$_2$O, are observed.
In AIMD simulations, quantum effects of nuclei can be included through the Feynman path-integral (PI) method \cite{feynman_quantum_1965}.
Recent PI-AIMD simulations \cite{morrone_nuclear_2008,ceriotti_nuclear_2013,ko_isotope_2019,cheng_ab_2019} show that NQEs altered the simulated structural properties by a considerable extent.
Moreover, as the zero-point energy enables atoms to explore classically inaccessible regions of the PES, the structural changes caused by NQEs highly depend on the description of the underlying PES, i.e., the choice of XC functionals.
Previous PI-AIMD studies employing various functionals \cite{marsalek_ab_2016,gasparotto_probing_2016,ko_isotope_2019,yao_temperature_2020,zhang_isotope_2020} show highly fluctuating results.
Yao \textit{et al.} \cite{yao_temperature_2020} studied the NQEs of liquid water using the SCAN functional in the canonical ensemble at different temperatures.
They showed that the NQEs brought the OH and HH RDFs much closer to the experimental values but the OO RDFs changed very little from its classical counterpart. 
Based on NQEs, a systematic study of isotope effects employing the state-of-the-art SCAN meta-GGA functional at ambient condition would be consequential on the path of the theoretical modeling of liquid water, which has not been fully addressed.

\par 
In PI-AIMD simulations, physical systems are represented by ring-polymers composed of several replicas (beads), which largely increase the computational costs.
Even with advanced methods \cite{ceriotti_nuclear_2009,ceriotti_efficient_2012} that reduce the required number of beads, the computational burden is still a vital issue that limits the application of PI-AIMD.
The rise of machine-learning-based approaches, e.g., the deep potential molecular dynamics (DPMD) method \cite{zhang_deep_2018}, has enabled overcoming this computational barrier.
With the DPMD method, one can train a deep neural network with a small amount of \textit{ab initio} data and perform extensive molecular dynamics simulations for an extended system with a linear-scaling computational cost which is orders of magnitude cheaper than traditional AIMD.
At the same time, DPMD is able to provide similar accuracy ($\sim$1 meV/molecule) as \textit{ab initio} approaches for water systems \cite{zhang_deep_2018}.
The DPMD model has been used to study the isotope effect in liquid water with the PBE0+TS XC functional \cite{ko_isotope_2019}, where quantitative agreements with experiment is achieved.

\par
The present work focuses on studying NQEs and isotope effects in liquid water via DPMD and PI-DPMD based on the PES provided by the SCAN XC functional within the isothermal–isobaric (N\textit{p}T) ensemble.
With the inclusion of NQEs, quantum fluctuations that can either strengthen or weaken the H-bond \cite{yao_temperature_2020,li_quantum_2011} are observed.
Remarkable changes are illustrated in structural properties of liquid water such as RDFs, statistics of H-bond information, oxygen-oxygen-oxygen ADF, and local tetrahedrality.
Besides, we present quantum effects on electronic properties such as the density of state and the distribution of dipole moments of liquid water.
Both structural and electronic properties obtained from our simulations agree better with the experimental observations as a result of NQEs.
Furthermore, isotope effects are well reproduced. 

\par
The rest of this paper is organized as follows. 
Computational details about the preparation of training data, the training process, and (PI-)DPMD calculations are described in Sec. \ref{method}. 
Simulation results including structural and electronic properties are analyzed and compared with experiment in Sec. \ref{results}. 
Finally, we conclude our work with a short summary in Sec. \ref{conclusion}.

\section{METHODS}
\label{method}
\par 
In order to generate an accurate neural network force field based on the SCAN functional, we adopted the active learning procedure called deep potential generator (DP-GEN) \cite{zhang_active_2019}.
The procedures are summarized as follows.
First, a PI-AIMD simulation is performed to provide initial training data set for the DPMD model.
Second, four independent DPMD models are trained based on the same training data set but different initialization of the deep potential parameters. 
Third, based on the trained DPMD models, eight separated PI-DPMD simulations are performed to explore the PES.
Configurations, where the four independent DPMD models give diverse predictions, are marked as unexplored configurations.
Fourth, we generate energies and forces from \textit{ab initio} calculations for the selected configurations.
Then, these newly generated data are added to the training data set and go back to the second step until the satisfactory accuracy is reached.
The final DPMD model is used to generate the DPMD and PI-DPMD trajectories of liquid water.

\par 
For generating initial training data set, 
a Born-Oppenheimer Feynman path-integral \cite{feynman_quantum_1965} AIMD simulation of a periodically replicated cubic box containing 64 H$_2$O molecules was performed within N\textit{p}T ensemble at ambient conditions (300 K and 1.0 bar).
The PES was calculated based on the SCAN meta-GGA functional \cite{sun_strongly_2015} with the Quantum ESPRESSO (QE) package \cite{giannozzi_quantum_2009,giannozzi_advanced_2017}.
Hamann-Schl\"uter-Chiang-Vanderbilt (HSCV) norm-conserving pseudopotentials \cite{vanderbilt_optimally_1985} were used to model the core level electrons.
Valence electrons were represented by a plane-wave basis (with a 130 Ry effective cutoff and a 12.66 \r{A} reference cell \cite{chen_ab_2017}) as implemented in the QE package.
Only the gamma point was used to sample the Brillouin zone of the supercell.
In PI-AIMD, the integration time step was set to 0.48 fs.
An 8-bead ring-polymer with a colored-noise generalized Langevin equation thermostat \cite{ceriotti_nuclear_2009,ceriotti_efficient_2012} was adopted to model quantum nuclei using the i-PI package \cite{kapil_i-pi_2019}.
To expand the region that light protons are able to explore on the PES, light water was adopted in the simulation.
The simulation cell was propagated using the Parrinello-Rahman method \cite{parrinello_crystal_1980} and thermostatted by another colored noise Langevin thermostat with a fictitious mass consistent with a 200 fs timescale. 
The total length of our final PI-AIMD production run is over 10 ps.
This provides a rough exploration for the PES of the SCAN functional.

\par
Atomic coordination, energy ($E$), ionic force ($\bm F_i$), and stress tensor ($\bm\Xi$) were extracted from the PI-AIMD trajectory as the initial training data set.
These data were used as the input PES data to train four independent DPMD models using the DeePMD-kit package \cite{wang_deepmd-kit:_2018} as described in Ref. \onlinecite{zhang_deep_2018}.
First, the input data were transformed to a local coordinate frame \{$D_{ij}$\} for each atom and its neighbors within 6 \r{A}, in order to preserve the translational, rotational, and permutational symmetries of the system.
Then \{$D_{ij}$\} was used as input of a deep neural network (DNN), which includes five hidden layers with 240, 120, 60, 30, 10 nodes and returns the energy of atom $i$.
During the training process, the Adam method \cite{kingma_adam:_2014} was used to optimize the DNN parameters with a set of loss functions defined as:
\begin{equation}
    L(p_\epsilon,p_f,p_\xi)=p_\epsilon\Delta\epsilon^2 + \frac{p_f}{3N}\sum_i|\Delta \bm F_i|^2 + \frac{p_\xi}{9}||\Delta\bm\xi||^2,
\end{equation}
where N stands for the number of atoms, $\epsilon=E/N$, $\bm\xi=\bm\Xi/N$, and $\Delta\epsilon$, $\Delta \bm{F}_i$, $\Delta\bm\xi$ denote the difference between the input data and current DPMD predictions.
$p_\epsilon$, $p_f$, $p_\xi$ are preset tunable parameters.
For efficiency, $p_\epsilon$, $p_f$, and $p_\xi$ were adjusted from 0.02 to 8, 1000 to 1, and 0.02 to 8, respectively, throughout the training procedure, while the learning rate decayed exponentially.
After training for $10^6$ steps, we obtained four DPMD models.
As the training data set may not cover the full PES, these four DPMD models will show different performances for unexplored configurations.

\par 
After obtaining the first generation of DPMD models, the DP-GEN iterations began.
To further explore the PES, eight additional PI-DPMD simulations were performed using the DeePMD-kit and the i-PI packages.
The driving force was provided by one of the four DPMD models.
We adopted exactly the same i-PI parameters as used in the generation of the initial training data set and initialize the simulations from independent starting configurations.
All trajectories lasted for around 1 ns.
Along the PI-DPMD trajectories, all four DPMD models were used to predict forces.
The differences between forces predicted by the four DPMD models were recorded for each configuration.
Configurations with maximum force differences greater than 0.2 eV/\r{A} were selected and marked as unexplored configurations.
If the proportion of marked configurations along the trajectory is less than 1\%, the four DPMD models were considered as converged.
Otherwise, we performed self-consistent \textit{ab initio} calculations using the QE package to calculate the ground state of the selected configurations.  
All related  parameters were kept the same as the initial PI-AIMD simulation.
\textit{Ab initio} energies and forces were extracted and added to the training data set.
Four new independent DPMD models were generated based on the updated training data set.
Then, we repeated the PES exploration step.
After 9 iterations, the DPMD model converged.
The final DPMD model was tested to have a root mean square error of 0.6 meV/atom for energy prediction.

\par 
With the DPMD model, the simulation size was expanded to include 128 water molecules in order to reduce the finite-size effects and optimize statistical sampling of the molecular dynamics trajectories.
Three simulations adopting the DPMD model were presented in this work.
First, a single classical DPMD simulation of liquid water within N\textit{p}T ensemble at ambient conditions was performed using the combination of the DeePMD-kit and the LAMMPS package \cite{plimpton_fast_1995}.
Nos\'e-hoover thermostat and barostat \cite{tuckerman_liouville-operator_2006} were used to control the temperature and pressure of the system.
The integration time step was set to 0.5 fs.
The resulting DPMD trajectory lasted for 5 ns.
The other two PI-DPMD simulations of liquid H$_2$O and D$_2$O were performed adopting similar parameters as the training process. 
For heavy water, the nuclear mass of hydrogen was set to 2.01 u.
The length of both PI-DPMD trajectories is 960 ps, which ensures that the computed properties are statistically converged.
Moreover, in order to study electronic properties, \textit{ab initio} level electron ground state as well as the Maximally localized Wannier Functions (MLWFs) \cite{marzari_maximally_1997,marzari_maximally_2012} centers were computed for 4000 selected snapshots uniformly distributed along each trajectories using the QE package. 
For the MLWFs calculations, we adopted a lower energy cutoff of 85 Ry and other simulation parameters are similar as used in the molecular dynamic calculations.

\section{Results and discussion}
\label{results}
\subsection{Proton transfer coordinates}

\par 
Quantum fluctuations result in two competing effects that either strengthen or weaken the H-bond in liquid water \cite{ceriotti_nuclear_2016}. 
The proton fluctuations along the stretching direction of O-H covalent bond can facilitate H-bond formations.
On the contrary, the fluctuations along proton libration direction, which are perpendicular to the H-bond, tend to weaken the H-bond network.
The former effects dominate in relatively strong hydrogen bonds, whereas the latter ones dominate in relatively weak hydrogen bonds \cite{li_quantum_2011}.
These two competing nuclear quantum effects are well captured by the SCAN functional as shown by the previous study \cite{yao_temperature_2020}.
The overall NQEs depend on the delicate balance between the two opposite effects. 

\par
\begin{figure}[ht]
\centering
\includegraphics[width=0.99\columnwidth]{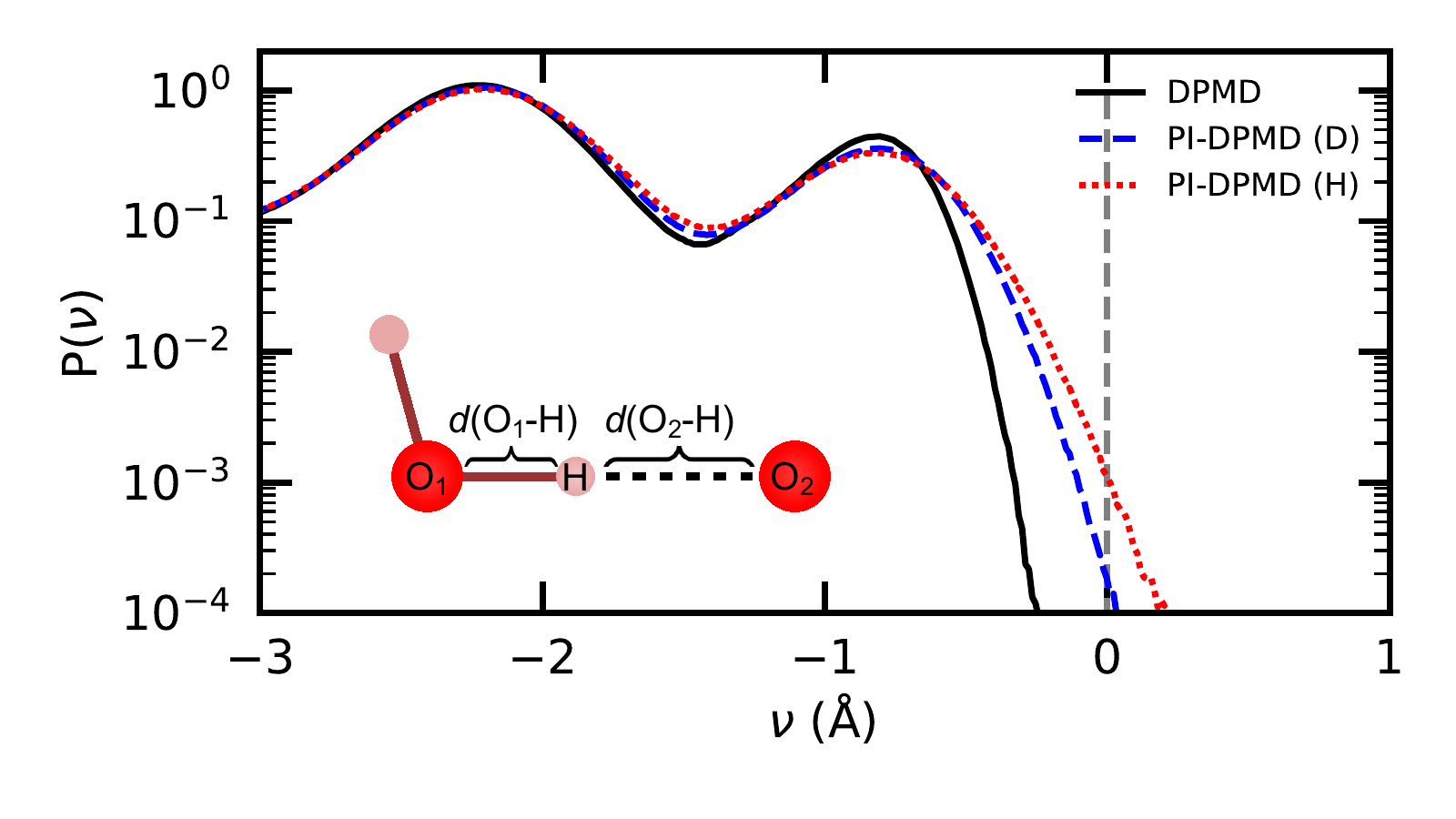}
\caption{
Log scale distribution of proton transfer coordinate $\nu$ of liquid water obtained from DPMD (black solid line), D$_2$O PI-DPMD (blue dashed line) and H$_2$O PI-DPMD (red dotted line) simulations. The inset shows the definition of $\nu=d(\text{O}_1\text{-H})-d(\text{O}_2\text{-H})$. 
}
\label{ptc}
\end{figure}

 
To illustrate fluctuations of protons under NQEs, we resort to the proton transfer coordinates $\nu$, which is defined as $\nu=d(\text{O}_1\text{-H})-d(\text{O}_2\text{-H})$, where $\text{O}_1$ and H denote the oxygen (O) atom and any of the two hydrogen (H) atoms in one water molecule and $\text{O}_2$ represents other oxygen atoms within the first coordination shell of the water molecule, as shown in the inset of Fig. \ref{ptc} \cite{ceriotti_nuclear_2013,sun_electron-hole_2018}.
The cutoff distances for defining the first coordination shell of a water molecular are chosen as the position of the first minimum of O-O RDF, which are listed in Table \ref{gr_peak}.
As defined, $\nu$ characterizes the location of H atoms between the two O atoms within two adjacent water molecules.
For example, $\nu = 0$ means that the H atom is precisely in the middle of two O atoms.
This configuration forms a strong H-bond and shows a high tendency to facilitate a proton transfer event.
The distribution of proton transfer coordinate $\nu$ generally exhibits two peaks.
The peak at lower $\nu$ ($< -1.3$) originates from H atoms that have $\text{O}_1$-H covalent bond pointed away from $\text{O}_2$.
The other peak ($\nu > -1.3$) comes from the H atoms located in between $\text{O}_1$ and $\text{O}_2$, which are likely to form H-bond. 

\par 
The proton delocalization along the direction of the acceptor oxygen, which induces stronger hydrogen bonds, can be seen by the wider distribution of $\nu$.
Fig. \ref{ptc} shows that the distributions of $\nu$ become broader from classical DPMD to D$_2$O and H$_2$O PI-DPMD simulations.
Moreover, a small amount of positive $\nu$ appears in both the PI-DPMD simulations, while only negative $\nu$ can be found in classical DPMD simulations.
The positive $\nu$ is more pronounced in the light water than the heavy water.
This is consistent with the finding in previous works \cite{ceriotti_nuclear_2013,sun_electron-hole_2018}.
Under NQEs, the zero-point energy (ZPE) enables the proton to overcome potential barriers and explore configurations that are hard to access in classical simulations.
As H$_2$O has larger ZPE than D$_2$O, light water shows over an order of magnitude more positive $\nu$ compared to heavy water in PI-DPMD simulations.
It is worth mentioning that these extreme H-bond fluctuations only last on a time scale of femtoseconds and no proton transfer event is observed on the time scale of the entire simulation; a similar observation that has been reported by Ceriotti \textit{et al.} \cite{ceriotti_nuclear_2013}.

\par
The proton delocalization along the libration direction, which weakens the hydrogen bonds, is revealed by the height of the H-bond peak and the position of the minimum between two peaks in Fig. \ref{ptc}.
In quantum simulations, the height of the H-bond peak is reduced from its classical counterpart by a larger extent than that of the none-bonded peak.
Also, the position of the minimum between two peaks shifts to the right by about 0.04 \r{A}.
The isotope substitution from D$_2$O to H$_2$O shows a similar trend with less extent in regard to the peak changes. 
These facts indicate that more water molecules originally H-bonded to each other in classical simulations become non-bonded under the libration fluctuations caused by NQEs.

\par
As a summary, both kinds of quantum fluctuations, that can either facilitate or weaken the H-bond network in liquid water, have been observed through the comparison of PI-DPMD simulations with DPMD simulations.
In the following sections, more properties will be discussed to determine how the above mentioned NQEs change the overall structure of liquid water.

\subsection{Radial distribution functions}

\par
\begin{figure}[ht]
\centering
\includegraphics[width=0.99\columnwidth]{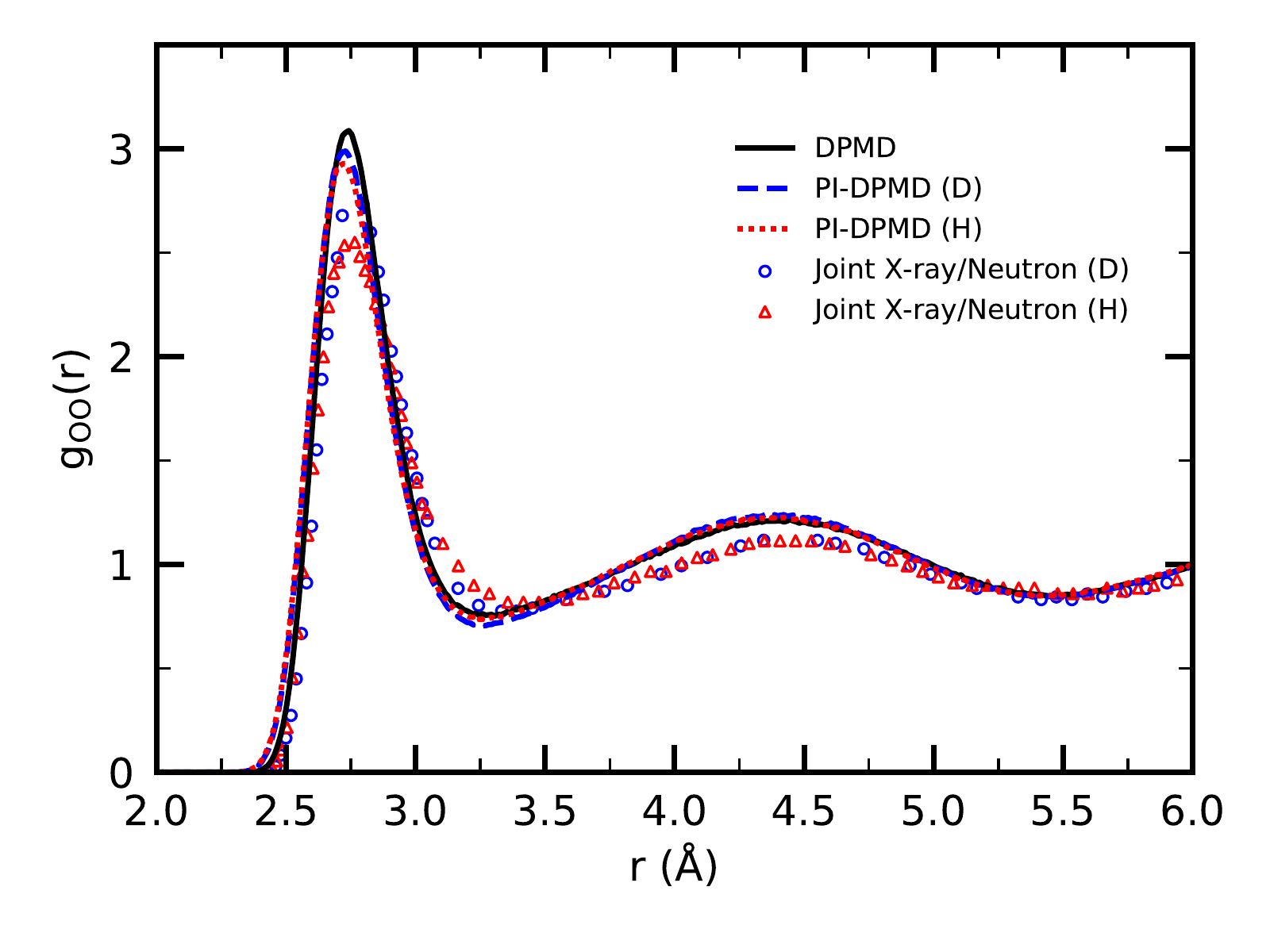}
\caption{
O-O RDFs of liquid water obtained from DPMD (black solid line), D$_2$O PI-DPMD (blue dashed line) and H$_2$O PI-DPMD (red dotted line). The  heavy water (blue circle) and light water (red triangle) joint x-ray/neutron experimental data are extracted from Ref \onlinecite{soper_quantum_2008}.}
\label{goo}
\end{figure}

\begin{figure}[ht]
\centering
\includegraphics[width=0.99\columnwidth]{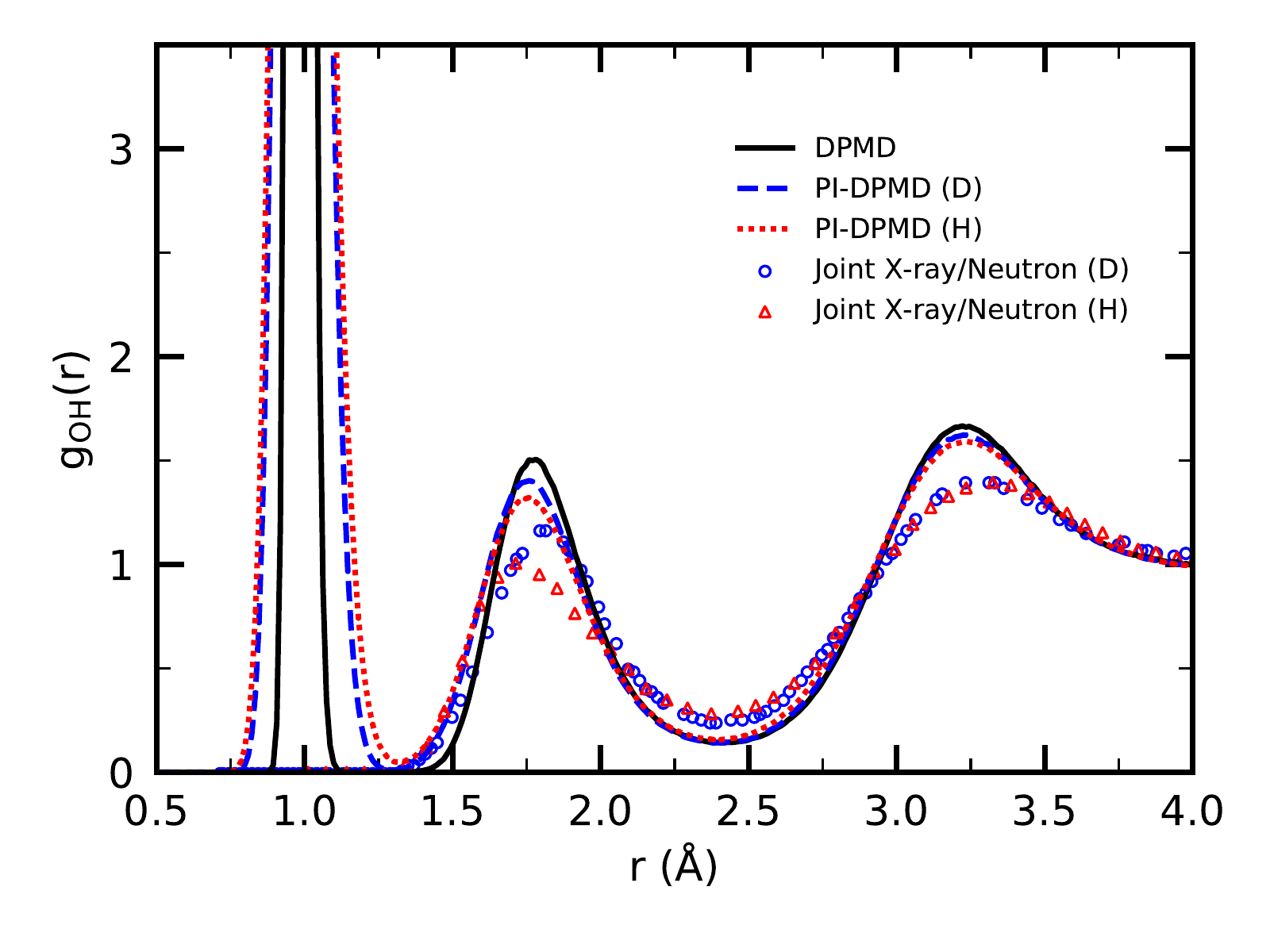}
\caption{
O-H RDFs of liquid water obtained from DPMD (black solid line), D$_2$O PI-DPMD (blue dashed line) and H$_2$O PI-DPMD (red dotted line). The  heavy water (blue circle) and light water (red triangle) joint x-ray/neutron experimental data are extracted from Ref \onlinecite{soper_quantum_2008}, which does not present the first peak around 1 \r{A}.}
\label{goh}
\end{figure}

\begin{figure}[ht]
\centering
\includegraphics[width=0.99\columnwidth]{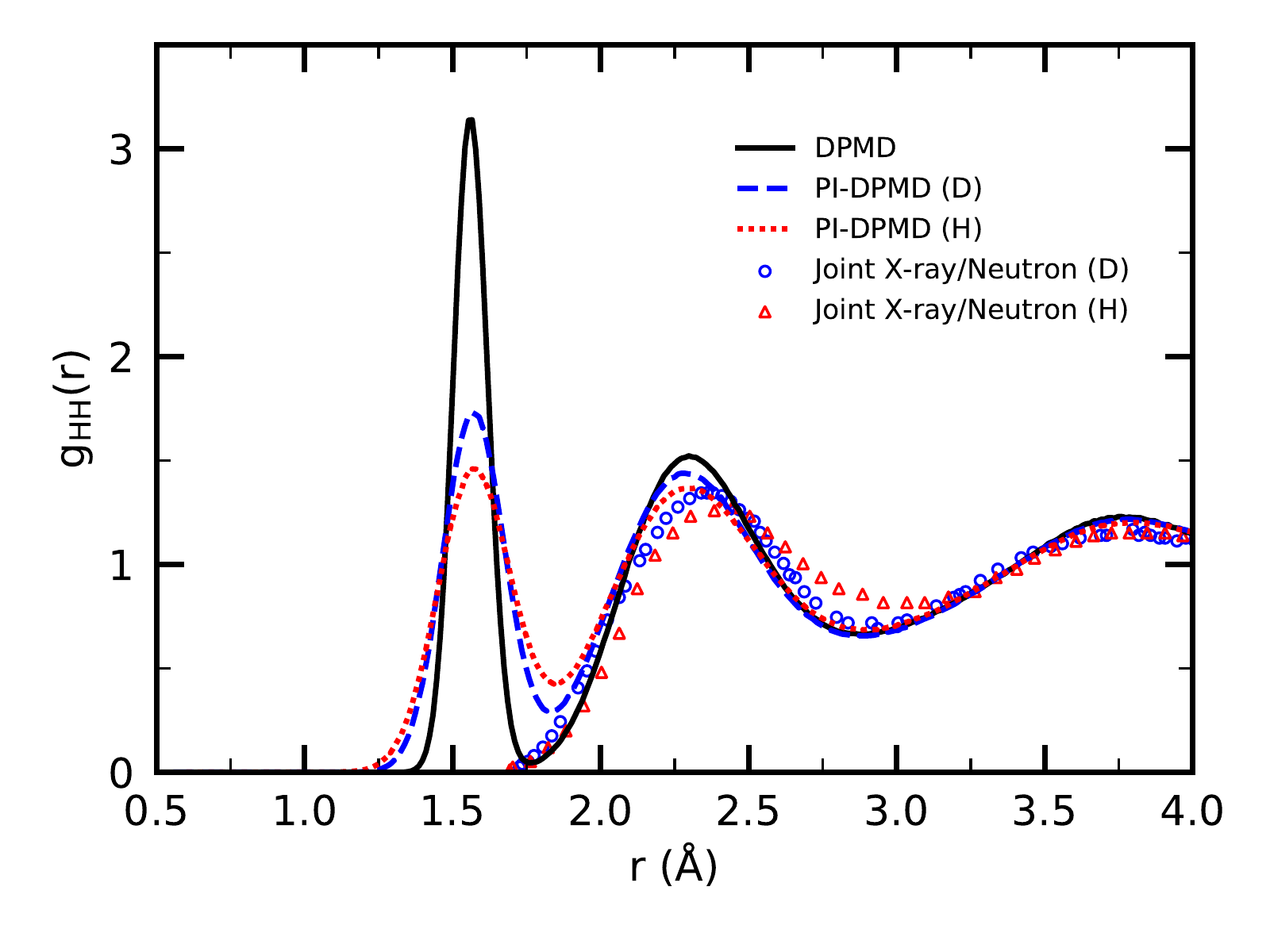}
\caption{
H-H RDFs of liquid water obtained from DPMD (black solid line), D$_2$O PI-DPMD (blue dashed line) and H$_2$O PI-DPMD (red dotted line). The heavy water (blue circle) and light water (red triangle) joint x-ray/neutron experimental data are extracted from Ref \onlinecite{soper_quantum_2008}, which does not present the first peak around 1.5 \r{A}.}
\label{ghh}
\end{figure}

 
Radial distribution functions are widely adopted tools to study the structure of water, which show the probability to find a given atom pair as a function of distance in real space.
In this section, O-O, O-H, and H-H RDFs from DPMD and PI-DPMD trajectories are shown in Figs. \ref{goo}, \ref{goh}, and \ref{ghh}.
For comparison, the EPSR data based on joint x-ray and neutron diffraction experiments by Soper \textit{et al.} \cite{soper_quantum_2008} are plotted as well.
The positions of maxima and minima of \textit{g}$_{\text{OO}}$(\textit{r}), \textit{g}$_{\text{OH}}$(\textit{r}), and \textit{g}$_{\text{HH}}$(\textit{r}) obtained from the simulations and experimental measurements are summarized in Table \ref{gr_peak}.
In general, one observes broadening effects in all O-O, O-H, and H-H RDFs when NQEs are included through the path-integral method. 
PI-DPMD simulations slightly soften water structures and yield better agreements with the experimental data as compared to the classical DPMD simulation.
Moreover, the RDFs show that D$_2$O is slightly more structured than H$_2$O due to the weaker zero-point motions.
The differences between the RDF of H$_2$O and that of D$_2$O qualitatively agree with experimental results, indicating that the experimentally observed delicate isotope effects are reproduced by our PI-DPMD simulations.

\begin{table*}[t]
\centering
\caption{Tabulated summaries of positions of maxima, minima for \textit{g}$_{\text{OO}}$(\textit{r}) , \textit{g}$_{\text{OH}}$(\textit{r}) and \textit{g}$_{\text{HH}}$(\textit{r}) obtained from classical DPMD, D$_2$O PI-DPMD and H$_2$O PI-DPMD.
The experiment data are extracted from Ref. \onlinecite{soper_quantum_2008} }
\label{gr_peak}
\begin{tabularx}{\textwidth}{>{\centering}X>{\centering}X>{\centering}X>{\centering}X>{\centering}X>{\centering}X>{\centering}X>{\centering\arraybackslash}X}
	\hline
	\hline
	\multirow{2}{*}{System}& \multicolumn{3}{c}{g$_{\text{OO}}$} & \multicolumn{2}{c}{g$_{\text{OH}}$} & \multicolumn{2}{c}{g$_{\text{HH}}$}\\ 
	\cline{2-4}\cline{5-6}\cline{7-8}
	&$r_{\text{max1}}$(\r{A}) &$r_{\text{min1}}$(\r{A}) &$r_{\text{max2}}$(\r{A}) &$r_{\text{max1}}$(\r{A}) &$r_{\text{max2}}$(\r{A}) &$r_{\text{max1}}$(\r{A}) &$r_{\text{max2}}$(\r{A})\\ 
	\hline
    DPMD&2.73(9)&3.27(5)&4.40(4)&0.98(0)&1.77(3)&1.56(0)&2.30(0)\\
    PI-DPMD (D$_2$O)&2.72(5)&3.26(0)&4.38(9)&0.98(2)&1.75(5)&1.57(1)&2.28(8)\\
    PI-DPMD (H$_2$O)&2.72(1)&3.26(5)&4.38(0)&0.97(9)&1.75(4)&1.57(4)&2.29(6)\\
    Expr. (D$_2$O)&2.76&3.35&4.45&&1.81&&2.37\\
    Expr. (H$_2$O)&2.75&3.40&4.44&&1.74&&2.42\\
	\hline
	\hline
\end{tabularx}
\end{table*}

\par 
Figure \ref{goo} shows the O-O RDFs, \textit{g}$_{\text{OO}}$(\textit{r}), which reflect the overall structures of liquid water.
The \textit{g}$_{\text{OO}}$(\textit{r}) obtained by PI-DPMD agrees better with the EPSR predictions as compared to DPMD results. 
The classical DPMD trajectory has a relatively more rigid first coordination shell and exhibits an over-structured first peak as compared to the PI-DPMD trajectories of light and heavy water.
Specifically, the height of the first maximum of \textit{g}$_{\text{OO}}$(\textit{r}) goes from 3.08 in DPMD down to 2.99 in heavy water and 2.93 in light water.  
On the other hand, NQEs slightly strengthen the long range correlations between water molecules.
The second peak is brought up by around 3\%, while the first minimum is brought down by around 5\% from DPMD to PI-DPMD. 
Therefore, the aforementioned two competing NQEs are well reflected in \textit{g}$_{\text{OO}}$(\textit{r}).
The above observations show that water structures go through minor but complicated overall structure changes under NQEs which agrees well with the conclusion of Yao \textit{et al.} \cite{yao_temperature_2020}.
In terms of isotope effects, heavy water shows a slightly higher first peak (by $\sim$0.06) than that of light water in PI-DPMD simulations; a result qualitatively agrees with the EPSR predictions (from 2.84 to 2.59).
Moreover, the experimentally predicted $\sim$0.01 \r{A} shift of the first \textit{g}$_{\text{OO}}$(\textit{r}) peak position upon isotope substitution is quantitatively reproduced. 
From D$_2$O to H$_2$O, the broadening of the first peak and slight height (from 0.71 to 0.74) and position (from 3.26 \r{A} to 3.27 \r{A}) increase of the first minimum can also be identified in PI-DPMD.
This indicates H$_2$O has slightly more non-bonded water molecules and shows softer structures as compared to D$_2$O.
As a result, the configurations of light water become more compact, which leads to a larger number density 0.10679 (0.10007) atom/\r{A}$^3$ than that of D$_2$O 0.10619 (0.10000) atom/\r{A}$^3$ in PI-DPMD simulations, where the number in parentheses denote the experiment value \cite{soper_quantum_2008}.
Furthermore, in both PI-DPMD simulation and the referenced experimental data, substituting D by H shows negligible effects on \textit{g}$_{\text{OO}}$(\textit{r}) beyond the second peak ($r>4.5$ \r{A}).
This indicates NQEs have insignificant impact on the long-range ordering of the H-bond network. 

\par 
The O-H RDFs, \textit{g}$_{\text{OH}}$(\textit{r}), are shown in Fig. \ref{goh}, where a similar quantum nuclei induced systematic softening effect can be distinguished.
The first peak of \textit{g}$_{\text{OH}}$(\textit{r}) corresponds to the O-H covalent bond, which is sensitive to the aforementioned quantum fluctuation along the stretching directions.
As listed in Table \ref{gr_peak}, peak positions are almost the same at around 0.98 \r{A} in all simulations.
However, as shown in Fig. \ref{goh}, the first peak exhibits an asymmetric distribution, where more H atoms tend to locate at the right side of the peak and form longer O-H covalent bonds, especially in the PI-DPMD simulations.
This can be understood by that the H atom experiences an asymmetric potential, where the potential barrier at the covalent bond side is steeper than that at the H-bond side.
Quantum fluctuations enables the H atom to explore more towards the less steep potential barrier direction, i.e. the H-bond side.
Thus, the peak position is not a good indicator of the averaged O-H covalent bond length.
In our simulations the averaged O-H covalent bond length for classical DPMD, heavy water, and light water are calculated to be 0.985, 1.000, and 1.006 \r{A}, respectively.
The 1.5\% O-H covalent bond elongation from DPMD to PI-DPMD confirms the proton delocalization effect under NQEs.
In PI-DPMD simulations, H$_2$O shows a $\sim$0.5\% longer covalent bond length than D$_2$O, which is relatively small as compared to the 3\% elongation from the EPSR results.
However, our PI-DPMD simulations yields similar result to the 0.4\% O-H and O-D covalent bond difference in gas phase measured by infrared spectroscopy \cite{cook_molecular_1974} as well as a $\sim$0.5\% difference of O-H and O-D covalent bond predicted by Zeidler \textit{et al.} via neutron scattering data applying different analytical techniques \cite{zeidler_oxygen_2011}.
The second peak of \textit{g}$_{\text{OH}}$(\textit{r}) reveals the distribution of H-bonds in liquid water.
Similar to the covalent peak, the position of the second maximum of \textit{g}$_{\text{OH}}$(\textit{r}) is located at $\sim$1.76 \r{A} and barely changes in all simulations.
The integrated average H-bond length decrease from 1.88(1) in classical DPMD to 1.85(7) in D$_2$O PI-DPMD and 1.85(6) in H$_2$O.
The reduction in height of the H-bonding peak indicates the increased amount of breaking H-bonds in liquid water under NQEs.
Although the decrease of peak height from heavy water to light water are well reproduced in PI-DPMD simulations,
the large position shit of 0.07 \r{A} of the second peak that found in EPSR studies are not obtained in PI-DPMD simulations. 
The observed changes of the H-bond peak highly relate to the variation of local tetrahedral ordering in liquid water under NQEs which will be discussed in the next section.
As for the third peak of \textit{g}$_{\text{OH}}$(\textit{r}), the inclusion of NQEs slightly softens the structure and improves both the peak height and peak position towards the experimental directions.
The observed isotope effects beyond the second peak of \textit{g}$_{\text{OH}}$(\textit{r}) ($r>2.5$ \r{A}) are negligible in both PI-DPMD simulations and EPSR results.

\par The H-H RDFs, \textit{g}$_{\text{HH}}$(\textit{r}), are given in Fig. \ref{ghh}.
The first peak of \textit{g}$_{\text{HH}}$(\textit{r}) represents the distance between two H atoms in the same water molecule.
The peak positions increase from 1.56 \r{A} in classical DPMD to 1.57\r{A} in both PI-DPMD simulations.
However, the averaged H-O-H covalent bond angle are calculated to be the same (105.4 $^\circ$) in all simulations.
Thus, the increase of peak positions can be attributed to the elongation of O-H covalent bond length caused by NQEs.
Moreover, the first peak of \textit{g}$_{\text{HH}}$(\textit{r}) is broadened under NQEs.
This can be interpreted as that water molecules shows more spread configurations along both its libration and stretching directions.
With regard to the second and third peaks, the changes due to the inclusion of NQEs and isotope effects are similar as discussed before in the case of \textit{g}$_{\text{OO}}$(\textit{r}).
Quantum fluctuations slightly soften the liquid structures so that the curves obtained from PI-DPMD are flattened and fit the EPSR observations better compared with classical DPMD results.
Isotope effects on peak shapes, such as the asymmetric changes on two sides of the second peak, and peak positions are qualitatively captured but with less extent than the EPSR data.

\par 
In summary, both proton elongation along the O-H covalent bond stretching direction and the fluctuations along proton libration direction are confirmed in \textit{g}$_{\text{OO}}$(\textit{r}), \textit{g}$_{\text{OH}}$(\textit{r}), and \textit{g}$_{\text{HH}}$(\textit{r}) RDFs. 
NQEs induce minor but complicated structural changes in liquid water.
As to isotope effects, the overall changes in the RDFs indicate that the quantum fluctuations weakening H-bonds dominate over the quantum effects that facilitate H-bonds.
Therefore, the structure of light water becomes more disordered than heavy water.

\subsection{Statistics of H-bond ordering}

\begin{figure}[ht]
\centering
\includegraphics[width=0.99\columnwidth]{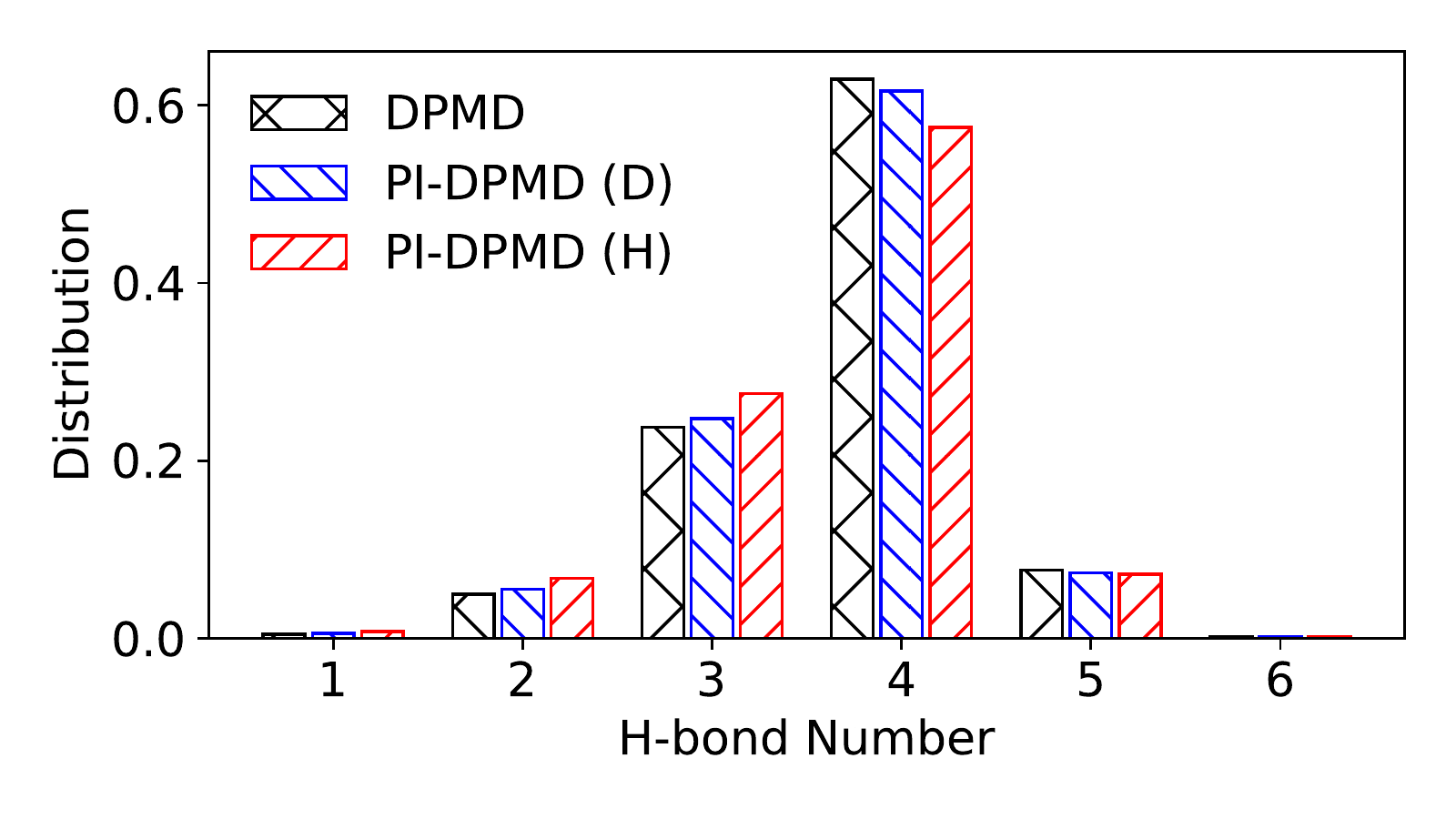}
\caption{
Distributions of average H-bond numbers per water molecule in liquid water obtained from DPMD (black solid), D$_2$O PI-DPMD (blue dashed) and H$_2$O PI-DPMD (red dotted).}
\label{hb}
\end{figure}

\par 
To further understand the local arrangement of H-bonds in water,  the average H-bond numbers formed by each water molecule are analyzed. 
We adopted the H-bond definition by Chandler \textit{et al.} \cite{luzar_hydrogen-bond_1996}, where two water molecules are considered H-bonded if the O-O distance is smaller than 3.5 \r{A} and O-O-H angle (the angle between O-O direction and O-H covalent bond direction) is less than 30$^{\circ}$. 
In an ideal tetrahedral structure, each water molecules accept two and donate two hydrogen atoms, forming four H-bonds.
In our DPMD, D$_2$O PI-DPMD, and H$_2$O PI-DPMD simulations the average H-bonds per molecule are 3.73, 3.70, and 3.64, respectively.
H-bond tends to break slightly more with the increase of quantum fluctuations.
To show the detailed information of the H-bonds, the distributions of averaged H-bond numbers are given in Fig. \ref{hb}.
Without NQEs, 63.0\% water molecules form standard tetrahedral structures, i.e. four H-bonds with its neighbouring water molecules.
This proportion goes down to 61.6\% and 57.5\% in PI-DPMD simulated heavy and light water.
Accordingly, the ratio of water molecules with less than four H-bonds rises in quantum simulations.
The isotope effect follows the same trend as the inclusion of NQEs.
This trend illustrates that with H-bond fluctuations under NQEs, some water molecules that form tetrahedral structures have one or more H-bond broken and squeeze into the interstitial region of other water molecules, leading to a weakened overall liquid structure.
Meanwhile, it is worth mentioning that other than a simple weakening effect of H-bond distribution, the proportion of water molecules with five H-bonds barely changes under NQEs.
It can be explained by that NQEs have different effects on H-bonds with different strength \cite{li_quantum_2011}. 
In other words, typical water-water H-bond are weakened but some strong ones are unchanged or even strengthened.
The above serves as another evidence of the fact that the result of NQEs is a subtle balance among different competing effects, which leads to overall weakened structures in liquid water.

\par 
To further explore the changes of H-bond arrangements upon isotope substitution, the three-body oxygen-oxygen-oxygen (O-O-O) ADF P$_{\text{OOO}}(\theta)$ and the tetrahedral order parameter $q$  of light and heavy water are studied and compared with EPSR results. 
The O-O-O ADF gives the probability of finding the triplet angle $\theta_{123}$ formed by oxygen atoms O$_1$, O$_2$, and O$_3$ in a local environment.
O$_1$ and O$_3$ are located within a cutoff distance $d$ from O$_2$, and $d$ is chosen to yield an average O-O coordination number of 4.0 \cite{soper_quantum_2008}.
In D$_2$O PI-DPMD and H$_2$O PI-DPMD, the values of $d$ are set to 3.205 and 3.210 \r{A}, respectively.
The tetrahedral order parameter $q$ is defined as
\begin{equation}
    q = 1 - \frac{3}{8}\langle\sum_{j=1}^3\sum_{k=j+1}^4(cos\theta_{jik}+\frac{1}{3})^2\rangle,
\end{equation}
where for one water molecule $i$ four nearest neighbors are considered and the bracket represents an average over all water molecules.
It is important to note that $q$ has a value between 0 and 1, where 1 represents the perfect tetrahedral structure and smaller values stand for distorted tetrahedral structures.

\begin{figure}[ht]
\centering
\includegraphics[width=0.99\columnwidth]{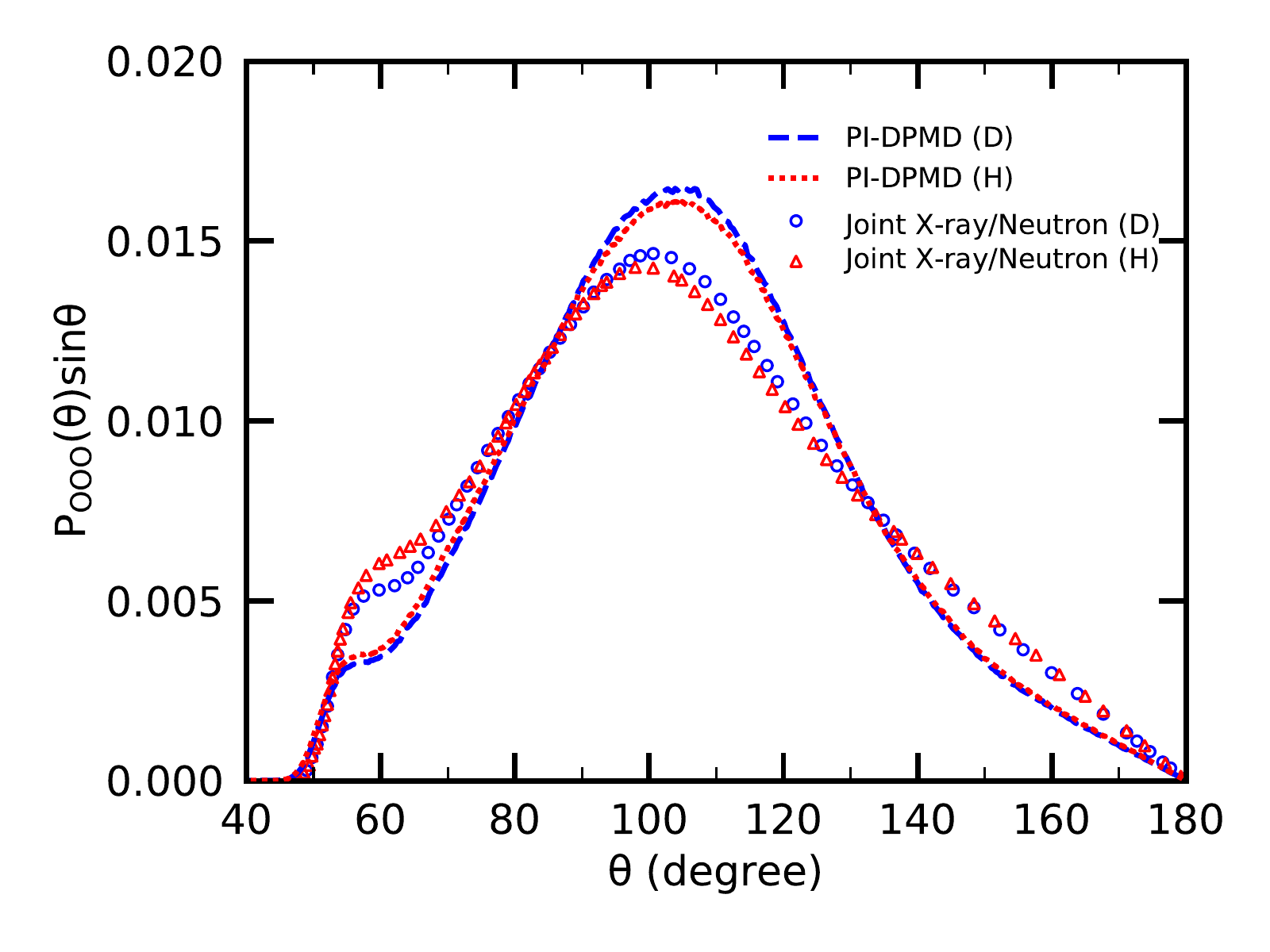}
\caption{
Triplet angular distribution of liquid water obtained from D$_2$O PI-DPMD (blue dashed line) and H$_2$O PI-DPMD (red dotted line) as well as the  heavy water (blue circle) and light water (red triangle) joint x-ray/neutron experimental data \cite{soper_quantum_2008}.}
\label{pooo}
\end{figure}

\par 
The triplet ADFs P$_{\text{OOO}}(\theta)$ are shown in Fig. \ref{pooo}.
Generally, P$_{\text{OOO}}(\theta)$ exhibits a principle maximum centered around 100$^\circ$ and a shoulder at around 60$^\circ$.
The main-peak angle around 100$^\circ$ is slightly smaller than the perfect tetrahedral angle 109.5$^\circ$, depicting a liquid structure with majority of water molecules forms tetrahedral structures with slight distortions. 
The shoulder arises as a result of water molecules in the interstitial regions that have partially broken H-bonds \cite{guillot_quantum_1998,chen_ab_2017}.
Isotope effects observed in the EPSR studies are qualitatively reproduced in our PI-AIMD simulations.
Going from heavy to light water, one notices that the main peak is broadened, the peak position is slightly shifted from to 104.8$^\circ$ and 104.0$^\circ$ and the peak height decreases by around 3\%.
Moreover, the side peak becomes more prominent in light water as compared to heavy water.
Both effects indicate the growth of the population of partially bonded water molecules, that donate either 0 or 1 H-bond to their neighbors and tend to form small-angle configurations, from D$_2$O to H$_2$O.
The changes of local structures under NQEs can be further confirmed by the tetrahedral order parameter.
From D$_2$O PI-DPMD to H$_2$O PI-DPMD, the values of $q$ changes from 0.714 (0.593) to 0.707 (0.576), with the values in the parenthesis representing the experiment/EPSR results.
Therefore, light water forms less local H-bonds and exhibits softer liquid structures.

\subsection{Electronic properties}

\par 
The strength of H-bond in liquid water is closely associated with the electronic properties of water molecules.
In this section, electronic properties such as density of states (DOS) of valence electrons and molecular dipole moments ($\mu$) are studied.
Even though \textit{ab-initio}-level electronic properties highly depend on the description of the underlying PES, which is determined here from the SCAN functional,
the resulting electronic structures are modified under NQEs due to the changes of characteristic geometry as described in previous sections.
Here we show that the PI-DPMD simulations improve the description of electronic properties in liquid water compared to the classical DPMD simulations.

\begin{figure}[ht]
\centering
\includegraphics[width=0.99\columnwidth]{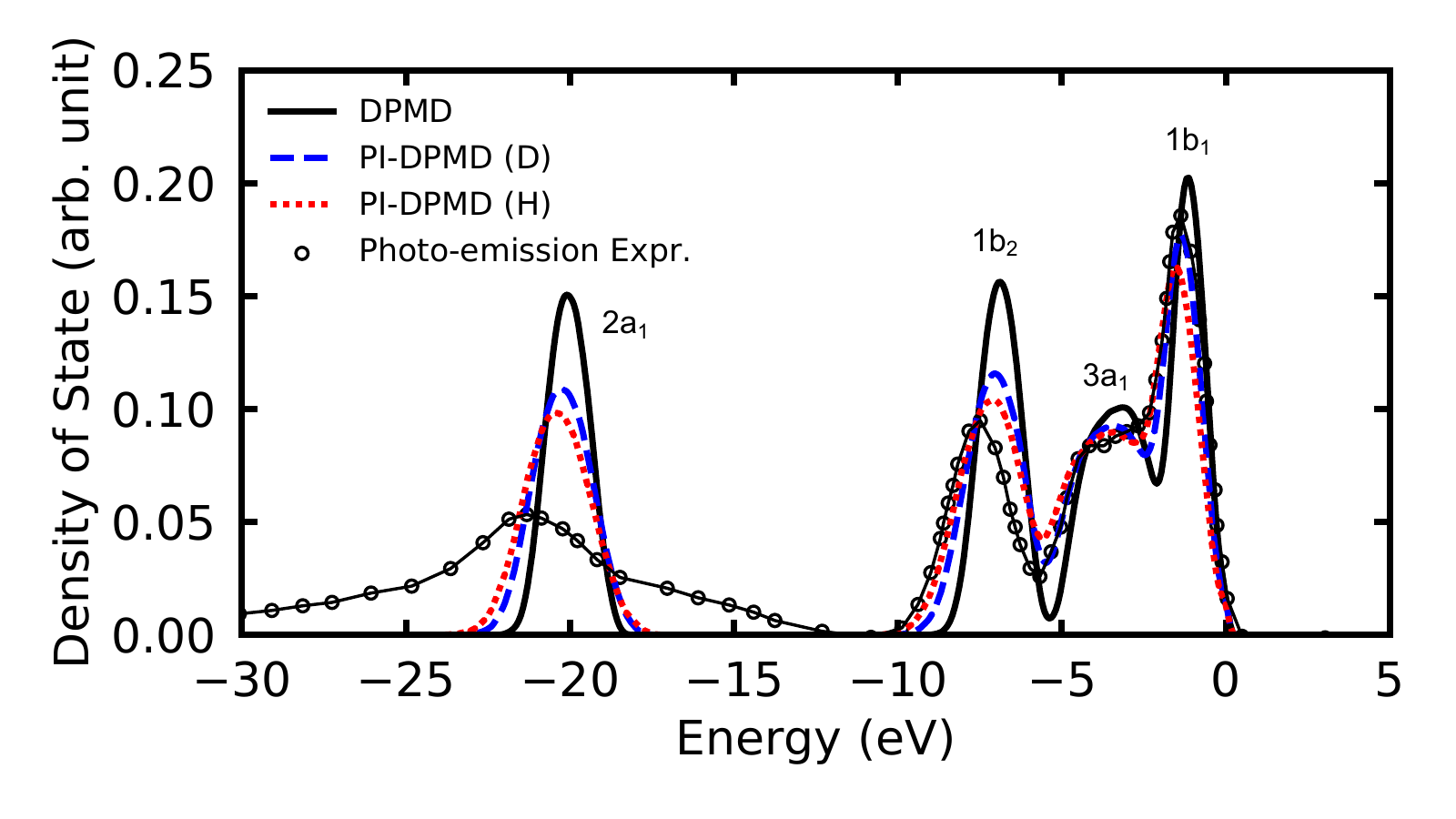}
\caption{
Density of state of liquid water obtained from DPMD (black solid), D$_2$O PI-DPMD (blue dashed) and H$_2$O PI-DPMD (red dotted). The density of state extracted from photoemission spectroscopy (black triangle) \cite{winter_full_2004} is plotted as a reference.}
\label{dos}
\end{figure}

\par
Figure \ref{dos} shows the DOS of liquid water obtained from the DPMD and PI-DPMD simulations as well as photoemission spectroscopy \cite{winter_full_2004}.
A 0.1 eV Gaussian broadening factor \cite{prendergast_electronic_2005} is adopted for the computed \textit{ab initio} eigenstates.
All DOS are aligned by the energy of the highest occupied states which is set to 0 eV.
The DOS exhibits 4 peaks, that can be assigned to the 2a$_1$, 1b$_2$, 3a$_1$, and 1b$_1$ valence orbitals of water.
These four orbitals are named by their spacial symmetries in the gas phase and labeled in Fig \ref{dos}.
On the one hand, as all our simulations are based on the SCAN DPMD model, the calculated binding energies of 2a$_1$, 1b$_2$, 3a$_1$ peaks with respect to the 1b$_1$ molecular orbitals are respectively around $-$2.0, $-$5.7, and $-$18.9 eV for all DPMD and PI-DPMD simulations, which is consistent with previous \textit{ab initio} studies adopting the SCAN functional \cite{chen_ab_2017,zheng_structural_2018} and fairly reproduces the result form the photoemission experiment ($-$2.3, $-$6.2, and $-$19.7 eV, respectively).
On the other hand, one finds the peak width of the DOS is sensitive to NQEs \cite{chen_ab_2016}.
In PI-DPMD simulations, all peak widths, especially the 2a$_1$ and 1b$_2$ orbitals that are largely associated with the covalent bond, get broadened by a large extent towards the experiment.
Also, the minima between peaks is better reproduced as well compared to experiments.

\begin{figure}[ht]
\centering
\includegraphics[width=0.99\columnwidth]{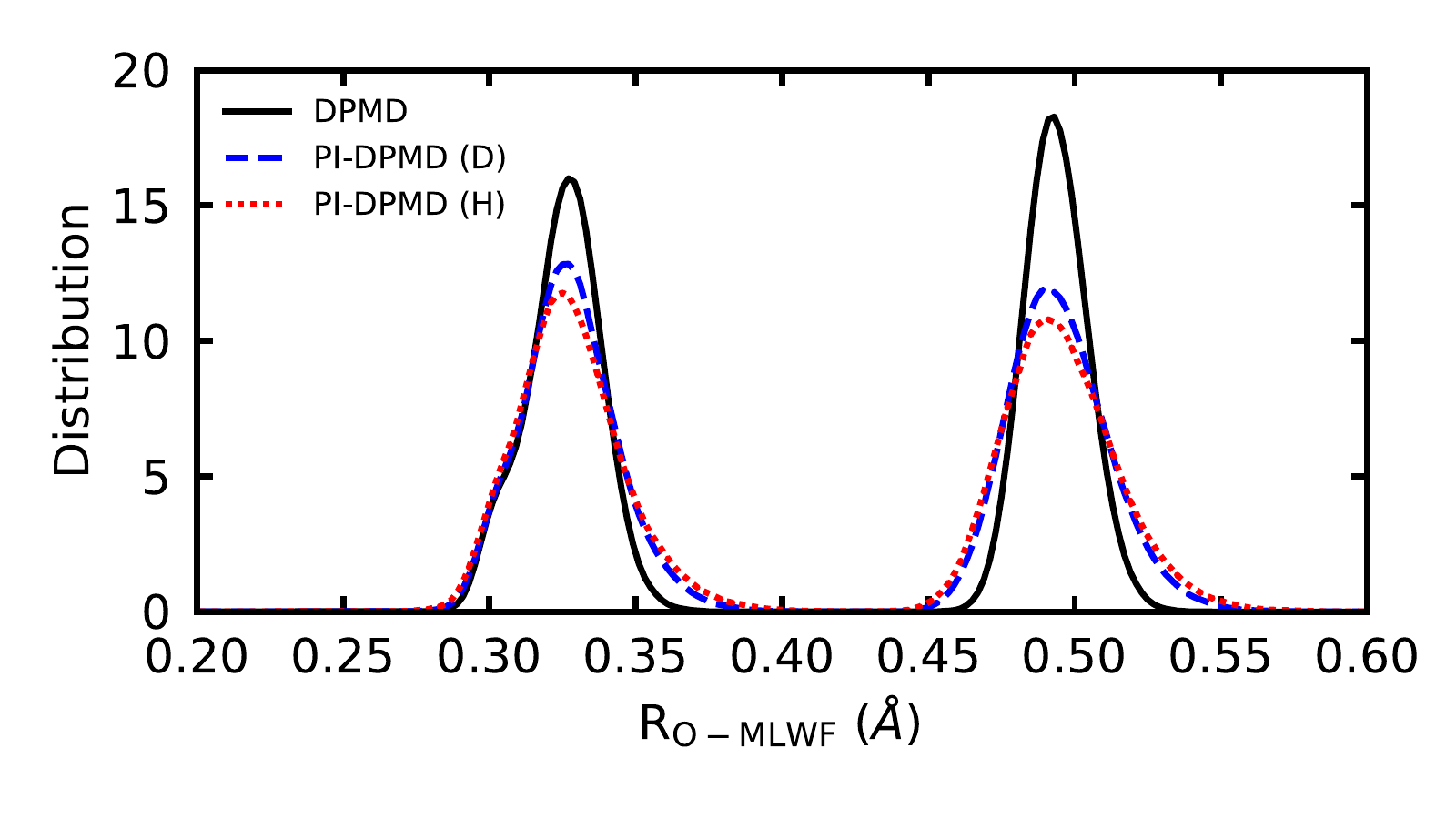}
\caption{
Distributions of the centers of MLWFs with respect to the oxygen position obtained from DPMD (black solid), D$_2$O PI-DPMD (blue dashed) and H$_2$O PI-DPMD (red dotted).}
\label{wannier}
\end{figure}

\par These four valance orbitals are correlated to four pairs of electrons in a water molecule.
Two of the electron pairs distribute along water covalent bond and are labeled as bonding pairs.
The other two electron pairs correspond to the lone pair of the oxygen atom.
These four electron pairs can be represented by a unitary transformation of the Kohn-Sham eigenstates with the spread minimized in real space, i.e. the MLWFs.
Based on the unitary transformation matrix, one finds the bonding pair electrons are composed of 1b$_2$, 2a$_1$ and 3a$_1$ orbitals (sorted by the proportion of contributions).
Similarly, the lone pair electrons are originated from 1b$_1$, 3a$_1$ and 2a$_1$ orbitals. 
The distributions of distances between the four MLWFs centers and the oxygen atoms are shown in Fig. \ref{wannier}.
The figure exhibits two peaks.
The peak located closer to the oxygen atom is originated from the lone pair electrons and the other peak represents the distribution of two bonding pairs.
Lone pair electrons are sensitive to the altering of the H-bond network, as the formation of H-bond directly relates to the attraction force between positively charged H atom and the lone pair electrons with negative charges.
In the classical simulation, the lone pair electrons show a subsidiary summit at 0.3 \r{A}.
This is attributed to the broken H-bonds in liquid water, where lone pair electrons experience less attractive force from the H-bond and move closer to the oxygen atom \cite{gaiduk_local_2017}.
Adding NQEs results in more broken H-bonds, as a result, the subsidiary summit is pushed to a slightly higher probability in both PI-DPMD simulations.
At the same time, more lone pair electrons move further away from oxygen to $\sim$0.37 \r{A}, which reveals the H-bond facilitated effect in NQEs.
In general, the lone pair peak moves from 0.328 \r{A} in DPMD to 0.326 \r{A} (heavy water) and 0.325 \r{A} (light water) in PI-DPMD.
The bonding pair peak locates at 0.492 \r{A} and barely shifts when NQEs are included, while the distribution of bonding pairs becomes more asymmetric and shows bias towards elongation.
These observations accord with the asymmetry of the O-H covalent bond distribution as described in \textit{g}$_{\text{OH}}$(\textit{r}).

\begin{figure}[ht]
\centering
\includegraphics[width=0.99\columnwidth]{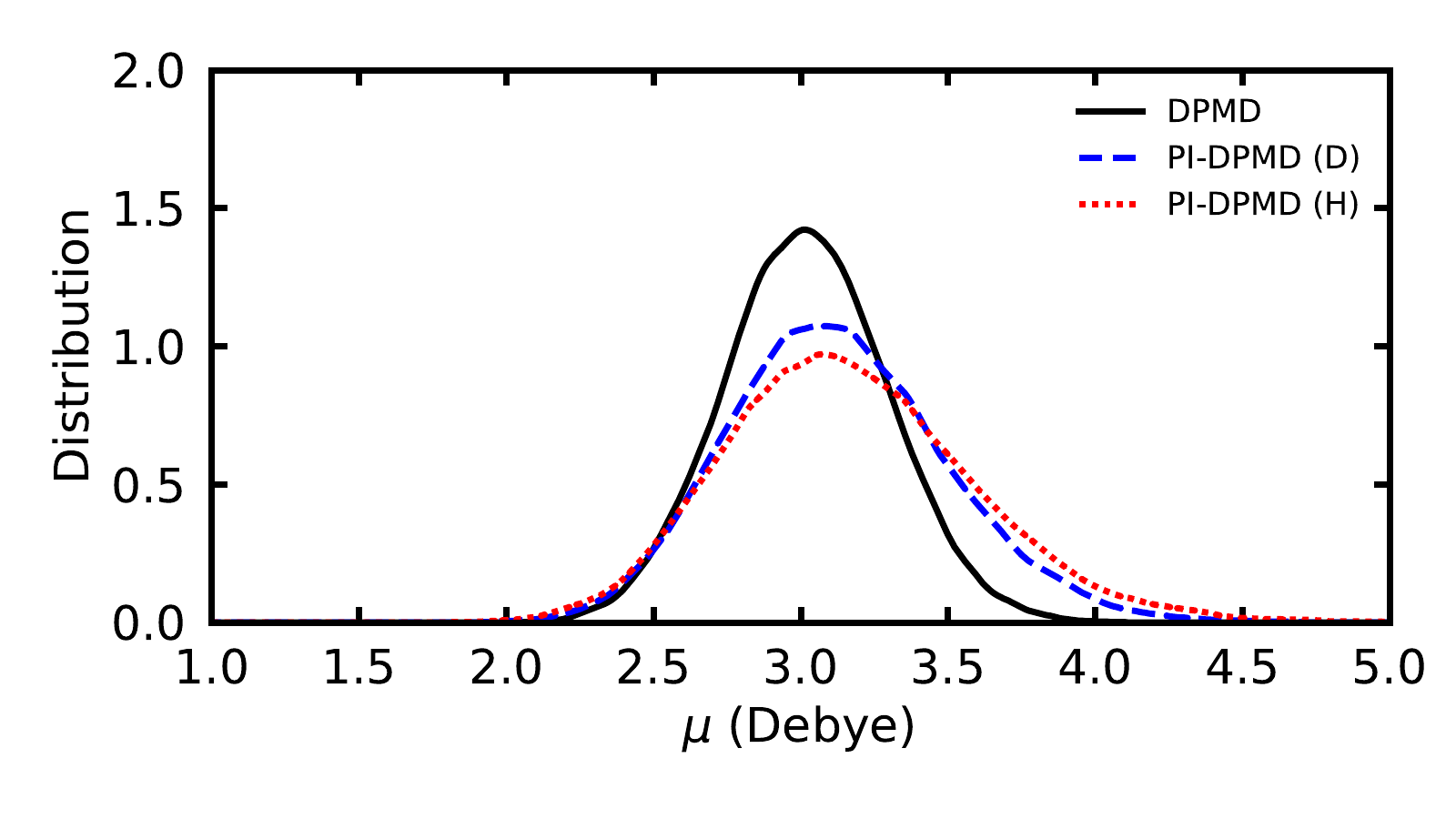}
\caption{
Distributions of dipole moments of water molecules obtained from DPMD (black solid), D$_2$O PI-DPMD (blue dashed) and H$_2$O PI-DPMD (red dotted).}
\label{dipole}
\end{figure}

\par The dipole moment of water molecules is an electronic property highly affected by the H-bond network in liquid water, which can be evaluated by the MLWFs center as: 
\begin{equation}
    \mu=\overrightarrow{\textrm R}_\textrm{OH$_1$} + \overrightarrow{\textrm R}_\textrm{OH$_2$} - 2\sum_{i=1}^4\overrightarrow{\textrm R}_\textrm{OW$_i$},
\end{equation}
where $\overrightarrow{\textrm R}_\textrm{OX}$ is the vector pointing from O atom to X which denotes either H atoms (H$_1$, H$_2$) or the $i$th MLWFs centers (W$_i$).
In the gas phase, the dipole moment of a single water molecule is measured to be 1.885 D \cite{chemical_rubber_company_crc_2003}.
Due to H-bonds, the value increases to 2.9 $\pm$ 0.6 D in liquid water \cite{badyal_electron_2000}.
The calculated average dipole moments in DPMD and D$_2$O, H$_2$O simulations are 3.01 $\pm$ 0.03, 3.12 $\pm$ 0.04, and 3.17 $\pm$ 0.05 D, respectively.
The polarization of water molecule increases with the growth of quantum fluctuations,
which is due to the delocalized protons under the NQEs effects.
As shown in Fig \ref{dipole}, a fraction of water molecules with extremely large dipole moment ($\mu > 3.5$ D) can be identified in PI-AIMD result.
Once again, this result fits the aforementioned observation of the O-H covalent bond elongation and the increased probability of proton transfers under NQEs.
However, the calculated dipole moments in our simulations are slighter larger than the experimental measurements.
One may expect that minimizing the self-interaction error by adopting higher level hybrid functionals would localize the distribution of electrons and reduce the calculated dipole moment of liquid water towards the experimental direction.

\section{CONCLUSION}
\label{conclusion}

In conclusion, we have systematically studied the NQEs and isotope effects in liquid water employing SCAN functional. 
Machine learning based DPMD and PI-DPMD simulations are performed at ambient conditions within the N\textit{p}T ensemble.
NQEs are shown to alter the structures of liquid water in different aspects.
Specifically, the quantum fluctuations can either strengthen or weaken the H-bond depending on the local configurations.
The predicted light water computed via SCAN PI-DPMD shows slightly softer structure as compared to the predicted heavy water. 
Both light and heavy water structures predicted by SCAN PI-DPMD are in good agreement with experiments.  
In particular, isotope effects on radial distribution functions, O-O-O triplet angular distribution as well as density found in experiment are qualitatively reproduced.
Descriptions of electronic structure by PI-DPMD are closer to experimental data than that of classical simulation.
Therefore, the combination of machine learning based model (DPMD), high-level \textit{ab initio} PES (SCAN) and quantum treatment of nuclei (PI) provides a promising approach for the modeling of liquid water with considerable time efficiency.
In future work, we expect to alleviate the remaining discrepancies by adopting a higher-level exchange-correlation functional, such as hybrid SCAN functional (SCAN0), to reduce the self-interaction error in DFT and provide more accurate structures for the study of isotope effect.

\begin{acknowledgments}
This work was supported by the Computational Chemical Center: Chemistry in Solution and at Interfaces funded by the DOE under Award No. DE-SC0019394.
B.S. was supported by the DOE BES under Grant No. DE-SC0018331.
The computational work used resources of the National Energy Research Scientific Computing Center (NERSC), a U.S. Department of Energy Office of Science User Facility operated under Contract No. DE-AC02-05CH11231. 
And this research includes calculations carried out on Temple University’s HPC resources and thus was supported in part by the National Science Foundation through major research instrumentation grant number 1625061 and by the US Army Research Laboratory under contract number W911NF-16-2-0189.

\end{acknowledgments}

\end{document}